\documentclass[galaxies,article,accept,pdftex,moreauthors]{Definitions/mdpi} 
\firstpage{1} 
\pubvolume{1}
\issuenum{1}
\articlenumber{0}
\pubyear{2026}
\copyrightyear{2026}
\externaleditor{Firstname Lastname} 
\datereceived{17 December 2025} 
\daterevised{5 March 2026} 
\dateaccepted{11 March 2026 } 
\datepublished{ } 

\Title{The MOND Depth Index and Dynamical Maturity Clock: Toward a Universal Classification of Galaxies and Star Clusters}

\Author{Robin 
 Eappen 
 $^{1,}$*\orcidA{} and Pavel Kroupa $^{1,2,}$*\orcidB{}}

\AuthorNames{Robin Eappen and Pavel Kroupa}

\address{%
$^{1}$ \quad Helmholtz-Institut für Strahlen- und Kernphysik (HISKP), Universität Bonn, Nussallee 14-16, \linebreak 53115 Bonn, Germany\\
$^{2}$ \quad Astronomical Institute, 
 Faculty of Mathematics and Physics, Charles University in Prague, V Holešovickách 2, CZ-180 00 Praha,  
 Czech Republic}

\corres{Correspondence: robin.eappen@gmail.com (R.E.); pkroupa@uni-bonn.de (P.K.)}

\abstract{Mass discrepancies in galaxies are empirically known to appear only below a characteristic acceleration scale $a_0$. Here we show that this behaviour is not limited to galaxies: it extends continuously across the full hierarchy of self-gravitating stellar systems, from gas-rich dwarfs and spirals to massive early-type galaxies, and further down to compact stellar clusters. We introduce the--- Milgromian dynamics (MOND) depth index $D_M$, together with dynamical maturity index $\mathcal{T}=t_{\rm cross}/t_H$, \textit{dynamical collisionality index} $\mathcal{T}_1=t_{\rm cross}/t_{\rm relax}$, with $t_{\rm cross}$ being the crossing time, $t_{\rm H}$ the Hubble time and $t_{\rm relax}$ the median two-body relaxation time, and the \textit{MOND acceleration index} $\mathcal{A}=\bar{a}/a_0$. {We uncover a well-defined two-dimensional dividing surface in dynamical space.} {The 'dark matter phenomenon' is found only in systems that are both in the deep-MOND regime ($\bar{a}<a_0$) and collisionless ($t_{\rm relax}>t_H$), while high-acceleration, collisional systems ($\bar{a}>a_0$, $t_{\rm relax}\ll t_H$), including globular clusters and UCDs, show no evidence for a mass discrepancy.} This clean dynamical separation defines a new, physically motivated classification scheme for stellar systems, unifying galaxies and clusters under one framework. {The observed division emerges naturally within the MOND framework and provides a useful diagnostic for examining how different gravitational paradigms account for the origin of the mass discrepancy.}}


\begin{document}
\section{Introduction}

The question of what fundamentally defines a galaxy continues to occupy a central place in modern astrophysics. Since the earliest attempts at classification, astronomers have sought a physical principle capable of organising the wide diversity of observed stellar systems. \citet{1919pcsd.book.....J,1928asco.book.....J} proposed evolutionary sequences of nebulae within the nebular-hypothesis framework, ideas that later influenced the conceptual basis from which the modern Hubble sequence emerged \citep{1926ApJ....64..321H}. Despite their historical importance, such morphological taxonomies remain largely descriptive and do not supply a single physical quantity that captures a galaxy’s structural maturity, dynamical state, or~stellar-population~age. 

A long-standing difficulty in galaxy studies is that many low-mass stellar systems occupy an ambiguous regime in which Newtonian gravity and baryonic matter alone cannot account for the observed kinematics, complicating attempts at a crisp operational definition of a “galaxy”. As~emphasised by \citet{2012AJ....144...76W}, a~physically meaningful definition should reflect whether a system’s dynamics require physics beyond baryons and Newtonian gravity. Several works have therefore argued for definitions based on the underlying dynamical state rather than morphology alone, for~example using relaxation time, size, or~stellar-population complexity as discriminants \citep{2011PASA...28...77F}. These approaches highlight that galaxies and star clusters form a continuum rather than two cleanly separated~classes.

More physically motivated definitions have focused on dynamical criteria. A~widely discussed operational distinction identifies galaxies as collisionless stellar systems with two-body relaxation times exceeding the Hubble time, as~emphasised by \citet{1998MNRAS.300..200K, 2008LNP...760..181K}. In~this picture, massive globular clusters and ultra-compact dwarfs (UCDs) lie close to the boundary between collisional and collisionless behaviour, whereas dwarf galaxies, spirals, and~ellipticals all behave effectively collisionlessly. The~degree of virialisation is likewise important: pressure-supported early-type galaxies (ETGs) generally lie close to virial equilibrium 
(\citet{1987gady.book.....B}), whereas rotationally supported disks may span a broader range of dynamical maturity. However, neither relaxation time nor virialisation uniquely maps onto stellar age, gas fraction, baryonic compactness, or~surface-brightness structure. Dwarf irregulars, low-surface-brightness (LSB) disks, high-surface-brightness (HSB) spirals, and~ETGs can exhibit overlapping dynamical timescales despite following distinct evolutionary~pathways.

\textls[-25]{Observational surveys have revealed strong empirical connections between galaxy mass, structural concentration, and~stellar-population age. The~downsizing phenomenon in which massive ETGs assemble early and quench rapidly is well} established (\citet{1996AJ....112..839C,2005ApJ...621..673T, 2009A&A...499..711R, 2010MNRAS.404.1775T, 2015MNRAS.448.3484M, 2019A&A...632A.110Y, 2022MNRAS.516.1081E, 2025NuPhB101716931G}). In~contrast, \textls[-25]{many gas-rich dwarf galaxies remain dynamically and chemically unevolved, exhibiting extended star-formation histories and predominantly young stellar populations} (\citet{Weisz2014, 2010ApJ...721..297M, 2024A&A...689A.221H}). These trends imply a deep connection between stellar age, surface density, and~baryonic structure. Yet, they remain fundamentally descriptive: no single scalar quantity predicts a system's position along this evolutionary continuum, and~conventional structural parameters (such as Sersic index, effective radius and rotational support) do not unify the full diversity of galaxy~types.

From a dynamical perspective, the~crossing time and characteristic acceleration provide partial insight. Diffuse galaxies with long crossing times, such as ultra-diffuse or LSB systems are expected to be less dynamically evolved than dense ETGs, which achieve virial equilibrium more rapidly (\citet{2015ApJ...798L..45V}). However, within~the standard $\Lambda$CDM framework, baryonic structure does not directly encode dynamical depth: halo concentration, feedback-driven expansion, angular-momentum exchange, and~merger history obscure any simple mapping~(\citet{2017ARA&A..55..343B}). {In standard dark-matter models, baryonic structure and total dynamical depth are not uniquely coupled, since baryons and dark matter are dynamically distinct components whose relation depends on assembly history and feedback.}

\textls[-15]{Modified Newtonian Dynamics (MOND), originally proposed by \citet{1983ApJ...270..365M}, offers a different perspective. MOND introduces a universal acceleration scale, }\mbox{$a_0 \approx 1.2\times10^{-10}\,\mathrm{m\,s^{-2}}$}, \textls[-15]{which underpins the baryonic Tully--Fisher relation} \citep{2000ApJ...533L..99M}, 
the radial-acceleration relation \citep{2017ApJ...836..152L}, 
\textls[-15]{and the tight correspondence between baryonic distributions and galaxy kinematics }
\citep{2012AJ....143...40M}. In~its non-relativistic formulation (\citet{1984ApJ...286....7B, 2023PhRvD.108h4005M}), MOND modifies the Poisson equation such that

\begin{equation}
    \nabla \cdot \left[ \mu\!\left( \frac{|\nabla\Phi|}{a_0} \right) \nabla\Phi \right] 
    = 4\pi G \rho,
\end{equation}
where $\Phi$ is the total gravitational potential, $\rho$ the baryonic density, and~$\mu(x)$ an interpolation function satisfying $\mu(x)\!\to\!1$ for $x\!\gg\!1$ (Newtonian regime) and $\mu(x)\!\to\!x$ for $x\!\ll\!1$ (deep-MOND regime), becoming the $p$ 
 = 3 Laplacian field equation, see \citet{2025A&A...698A.167S}. For~spherical systems this reduces to the algebraic MOND relation
\begin{equation}
    g\,\mu\!\left(\frac{g}{a_0}\right) = g_N,
\end{equation}
where $g$ is the true gravitational acceleration and $g_N=GM(r)/r^2$ is the Newtonian value. In~the deep-MOND limit ($g\!\ll\!a_0$), this yields the space-time scale-invariant form
\begin{equation}
    g \;=\; \sqrt{g_N\,a_0} \;=\; \sqrt{\frac{GM(r)\,a_0}{r^2}}.
\end{equation}

The appearance of the acceleration constant $a_0$ implies a characteristic MOND radius $r_M$ (discussed in Section~\ref{sec:theory}, Equation~(\ref{eq:rM}), {see \citet{1986ApJ...302..617M}}), beyond~which internal accelerations fall below $a_0$ and deep-MOND dynamics dominates. Defined purely by the baryonic mass and a universal constant, this radius delimits the onset of deep-MOND collapse behaviour: compact systems with most of their baryons interior to $r_M$ evolve rapidly and become dense and old, whereas diffuse systems with most baryons beyond $r_M$ evolve slowly and remain dynamically~young.

These considerations suggest the possibility of a single, physically motivated scalar quantity derived solely from the baryonic distribution relative to the MOND radius that could act as a structural or dynamical `clock' for galaxies. In~this work, we develop such an index, which we refer to as the MOND 
 depth index ($D_{M}$). While its formal definition is presented in Section~\ref{sec:theory}, the~basic concept is intuitive: the fraction of a system's baryons lying in the deep-MOND regime encodes its dynamical depth, collapse history, and~likely stellar age. Compact, high-acceleration systems are expected to lie predominantly within $r_M$ and correspond to old, virialised, quenched galaxies, while diffuse, low-acceleration systems reside deep in the MOND regime and correspond to younger, gas-rich, dynamically unevolved systems. Unlike in $\Lambda$CDM, MOND's universal acceleration scale makes this mapping physically meaningful and potentially~predictive.

The purpose of this paper is twofold. First, we introduce a set of physically motivated, MOND-based dynamical diagnostics—$D_M$, $\mathcal{T}$, $\mathcal{T}_1$ and  $\mathcal{A}$. Each of these depends only on the observed baryonic mass distribution and the characteristic MOND acceleration $a_0$. Second, we investigate whether these quantities collectively organise diverse stellar systems, including ETGs from the ATLAS$^{3\mathrm{D}}$ survey, HSB and LSB spirals, gas-rich dwarfs, and~compact clusters/UCDs, into~a continuous sequence of structural and dynamical maturity. By~comparing $D_M$, $\mathcal{T}$, $\mathcal{T}_1$ and $\mathcal{A}$ with stellar ages, gas fractions, internal accelerations, and~morphological class, we assess whether a unified, baryonic set of MOND-based diagnostics can reproduce the observed hierarchy of galaxy evolution and simultaneously provide a novel, purely empirical test of MOND~itself.

\section{Dynamical Framework and~Definitions}
\label{sec:theory}

To compare galaxies and stellar clusters within a unified dynamical context, we employ a set of physically motivated quantities that characterize internal timescales, gravitational depth in terms of the degree to which a system resides in the Newtonian or deep-MOND regime (Note: Dwarf satellite galaxies of the Milky Way are in the deep-MOND regime but are not isolated systems: they have typically lost their gas through environmental processes such as ram-pressure or tidal stripping. For~this reason, the~dynamical indices introduced here are primarily valid for non-satellite galaxies). All quantities are defined using only observable baryonic masses, characteristic radii, and~velocity scales, ensuring that the analysis remains~baryon-based.

The first fundamental quantity is the dynamical or crossing time, which measures the characteristic time for an object to traverse the system.  
For pressure-supported systems, we~adopt
\begin{equation}
    t_{\rm cross} = \frac{2R}{\sigma},
\end{equation}
where $R$ is a representative radius (typically the half-light or effective radius; see \mbox{\citet{1990ApJ...356..359H}},  \citet{2011MNRAS.413..813C}) and $\sigma$ is the line-of-sight velocity dispersion.  
For rotationally supported systems we analogously define
\begin{equation}
    t_{\rm cross} = \frac{\pi R}{V_{\rm c}},
\end{equation}
using the circular velocity $V_{\rm c}$ measured at a characteristic radius (e.g.,~\citet{2016AJ....152..157L}).  
Compact systems therefore exhibit short crossing times and rapid internal evolution, while diffuse galaxies have long dynamical times and evolve~slowly.

Another fundamental time scale characterizing gravitationally bound stellar systems is the classical median two-body relaxation time, which quantifies the timescale on which stellar encounters redistribute energy. For~a system of $N$ equal-mass stars with characteristic stellar mass $M_\star$ and half-mass radius $R_{\rm h}$, we adopt the standard approximation (\citet{2008gady.book.....B})
\begin{equation}
    t_{\rm relax} \approx \frac{0.1\,N}{\ln(0.4N)}\, t_{\rm cross}.
\end{equation}
Following 
 the arguments of \citet{1998MNRAS.300..200K, 2008MNRAS.386..864D} and with $t_{\rm H}$ being the Hubble time, systems with $t_{\rm relax} > t_H$ behave effectively as collisionless galaxies, while systems with $t_{\rm relax} < t_H$ are collisional and dynamically evolve as star clusters. Relaxation and virialisation must be clearly distinguished: a system may satisfy the virial theorem,
\begin{equation}
    2K + W = 0 ,
\end{equation}
after only a few crossing times, yet still possess a relaxation time far exceeding the Hubble time.  
Galaxies are therefore virialised but collisionless objects, whereas globular clusters and some UCDs can be both virialised and older than one median two-body relaxation time (\citet{1997A&ARv...8....1M}).

A central element of our analysis is the internal gravitational acceleration compared to the ubiquitous MOND acceleration constant \(a_0\) discovered by  \citet{1983ApJ...270..365M}. MOND introduces a characteristic scale that marks the transition between Newtonian and scale-invariant deep-MOND dynamics. For~a system of baryonic mass \(M_{\rm bar}\), one may define a MOND radius by equating the Newtonian acceleration \(GM_{\rm bar}/r^2\) with \(a_0\):
\begin{equation}
    r_M = \sqrt{\frac{G M_{\rm bar}}{a_0}}.
    \label{eq:rM}
\end{equation}

This radius, implicit in Milgrom's original formulation and made explicit in later MOND analyses (e.g.,~\ \citet{2012LRR....15...10F}), identifies the scale beyond which internal accelerations fall below \(a_0\) and deep-MOND behaviour dominates. Systems whose baryonic mass lies predominantly inside \(r_M\) are expected to undergo rapid early collapse and become dynamically compact and old, whereas systems with a large fraction of their baryons outside \(r_M\) evolve in the low-acceleration regime and therefore are expected to collapse more slowly (e.g.,~\ \citet{2008MNRAS.386.1588S, 2022MNRAS.517.3613K}).

To quantify how deeply a system resides in the MOND regime, we introduce the \textit{MOND depth index}
\begin{equation}
    D_M = 1 - \frac{M(<r_M)}{M_{\rm bar}},
    \label{Dm}
\end{equation}
where \(M(<r_M)\) is the baryonic mass enclosed within the MOND radius. In~this work, the~enclosed mass is computed from analytic profiles matched to each galaxy type: for ETGs we assume a Hernquist profile \citep{1990ApJ...356..359H} with scale length \(a = R_{\rm e}/1.8153\), where $R_{\rm e}$ is the effective radius, while for rotationally supported SPARC disks and dwarfs we adopt an exponential disk with scale length \(R_{\rm d}\). The~quantity \(D_M\) ranges from \(D_M \approx 0\) for compact, Newtonian-like systems to \(D_M \approx 1\) for diffuse, deep-MOND systems. Because~\(D_M\) depends solely on the baryonic mass distribution and the universal constant \(a_0\), it provides an observationally accessible measure of gravitational depth and structural maturity applicable across all galaxy types (e.g.,~\ \citet{2017ApJ...836..152L,2024A&A...689A.221H}).

A second key quantity is the dimensionless dynamical maturity index,
\begin{equation}
    \mathcal{T} = \frac{t_{\rm cross}}{t_H},
\end{equation}
which measures the number of internal dynamical cycles a system has experienced over the age of the Universe.  
Systems with $\mathcal{T} \ll 1$ have undergone many internal dynamical cycles and are dynamically old (e.g.,\ ETGs), while systems with $\mathcal{T}$ approaching unity have experienced only a few crossing times and remain dynamically young (e.g.,\ LSB disks and diffuse dwarfs). $\mathcal{T}$ is well defined for both pressure-supported and rotationally supported systems, making it particularly well suited to the broad comparative analysis pursued~here.

A long-standing dynamical criterion is that galaxies should behave as collisionless stellar systems, with~two-body relaxation times exceeding the Hubble time \citep{2011PASA...28...77F} (see 
 their discussion on using relaxation time as a galaxy-defining property). This motivates the inclusion of the dynamical collisionality index
\begin{equation}
    \mathcal{T}_1 = \frac{t_{\rm cross}}{t_{\rm relax}},
\end{equation}
which measures how many crossing times occur per relaxation time in our diagnostic framework. This ratio therefore provides a convenient, morphology independent description between galaxies (collisionless) and star clusters (collisional), complementing earlier work by \citet{1998MNRAS.300..200K,2008MNRAS.386..864D}.

Finally, to~characterise gravitational depth in a morphology-independent manner, we define a mean internal acceleration
\begin{equation}
    \bar{a} = \frac{G M_{\rm bar}}{R_e^2},
\end{equation}
and normalise it by the MOND acceleration constant to form another dimensionless index which we define as the acceleration index,
\begin{equation}
    \mathcal A = \frac{\bar{a}}{a_0} .
\end{equation}
Large values of $\mathcal A$ \textls[-25]{correspond to compact, dynamically deep systems that lie predominantly inside $r_M$, whereas small values identify diffuse systems operating well within the deep-MOND regime} (e.g., \citet{2013ApJ...775..139M, 2012LRR....15...10F, 2022MNRAS.517.3613K}).

Together, these indices $D_M$, $\mathcal{T}$, $\mathcal{T}_{1}$, and~$\mathcal A$ provide a coherent and fully baryonic set of dynamical descriptors applicable across the entire hierarchy of stellar systems. These indices form the basis for the diagnostic planes introduced in Section~\ref{sec:results}, where they reveal a continuous dynamical sequence connecting star clusters, dwarfs, spirals, and~ETGs.
\section{Data and Sample~Selection}
\label{sec:data}

To assess whether the $D_M$ provides a unified dynamical description across the full hierarchy of stellar systems, we assemble a heterogeneous but well characterized sample drawn from major observational surveys and literature catalogs. The~combined dataset spans ETGs, HSB and LSB spirals and dwarfs, compact stellar systems, including globular clusters and UCDs. The~inclusion of compact stellar systems is particularly important because UCDs and massive globular clusters extend smoothly into the parameter space of dwarf galaxies, forming a continuous size–mass sequence \citep{2011MNRAS.414.3699M}. This structural continuity reinforces the need for a unified dynamical diagnostic applicable across the full hierarchy of stellar systems. A~summary of the datasets and the key quantities used in our analysis is provided in Table~\ref{tab:datasets}. For~every system we compile baryonic masses, characteristic radii, velocity scales, and, where possible, stellar-population ages. These observables allow a homogeneous computation of the dynamical quantities defined in Section~\ref{sec:theory}, including crossing times, relaxation times, internal accelerations, MOND radii, and~the $D_M$.
\begin{table}[H]
\caption{Summary of the datasets used in this~study.}

\begin{adjustwidth}{-\extralength}{0cm}
\begin{tabularx}{\fulllength}{lccC}
\toprule
\textbf{Dataset/Type} &
\textbf{Mass} &
\textbf{Radius} &
\textbf{Age} \\
\midrule

ATLAS$^{3\!D}$ (ETGs) &
$M_\star$ (\citet{2011MNRAS.413..813C}) &
$R_{\rm e}$ (\citet{2011MNRAS.413..813C}) &
luminosity--weighted ages (\citet{2015MNRAS.448.3484M}) \\

SPARC (disks \& dwarfs) &
$M_\star$, $M_{\rm gas}$ (\citet{2016AJ....152..157L}) &
$R_{\rm d}$ (\citet{2016AJ....152..157L}) &
$t_{\rm gas}$ (Equation~(\ref{eq:tgas})) \\

MCO (GCs, UCDs, etc.) &
$M_{\rm dyn}$ (\citet{2008MNRAS.386..864D}) &
$R_{\rm h}$ &
old ($\geq$10--12 Gyr) \\
\bottomrule
\end{tabularx}
\end{adjustwidth}
\label{tab:datasets}
\noindent{\footnotesize{Note
. For ATLAS$^{3\!D}$ early-type galaxies we use stellar masses, effective radii, and~luminosity-weighted ages from the indicated references. For~SPARC disks and dwarfs we adopt stellar and gas masses (including helium), exponential scale lengths, and~gas-consumption ages $t_{\rm gas}$. For~MCOs we use dynamical masses and half-light radii from the literature. These quantities provide the inputs for computing the $D_M$, $\mathcal{T}$, $\mathcal{T}_1$ and $\mathcal{A}$ across all stellar~systems.}}
\end{table}
\textls[-15]{The ETG component is drawn primarily from the ATLAS\textsuperscript{3D} survey (\mbox{\citet{2011MNRAS.413..813C}}),} which provides homogeneous integral-field spectroscopy, photometry, and~dynamical modelling for 260 nearby ETGs. For~the present analysis, we select those ETGs with published luminosity-weighted stellar ages from \citet{2015MNRAS.448.3484M}, ensuring a consistent treatment of stellar populations. Effective radii, stellar masses, and~velocity dispersions are taken from \citet{2013MNRAS.432.1709C}, providing the necessary inputs for computing $t_{\rm cross}$, $\bar{a}$, and~$r_M$ under the baryonic framework adopted~here.

Rotationally supported spirals and dwarfs are taken from the SPARC database \citep{2016AJ....152..157L}, which provides high-quality rotation curves and H\,{\sc i} measurements for 175 galaxies covering the full range of disk morphologies and surface brightnesses, from~HSB spirals to extremely diffuse, gas-rich dwarfs. From~SPARC we adopt stellar masses, gas masses (multiplied by 1.33 to include helium; \citet{1997ApJ...481..689M}), exponential disk scale lengths, and~characteristic circular velocities $V_{\rm c}$.

We use the sample of Massive Compact Objects (MCOs) from Tables~1 and 2 of \citet{2008MNRAS.386..864D} (also referred to as D08), which includes globular clusters, UCDs, and~intermediate compact systems with published dynamical mass estimates, half-light radii, and~internal velocity dispersions. Following their classification, we treat all these objects uniformly as MCOs. Although~such systems are collisional on long timescales (\citet{1997A&ARv...8....1M}), their inclusion enables us to test whether $D_M$ can connect collisional star clusters and collisionless galaxies along a single continuous dynamical sequence as an extension of the traditional relaxation-based classifications (\citet{1998MNRAS.300..200K, 2011PASA...28...77F}). 

For the SPARC galaxies we estimate a simple gas-consumption age rather than adopting literature stellar ages, which are not homogeneously available for the full sample. Conceptually, a~gas-consumption timescale may be written as $t_{\rm gas} \approx M_{\rm gas}/{\rm SFR}$ (e.g. \citet{1998ApJ...498..541K}). In~practice, we lack SFR measurements for many SPARC galaxies, so we approximate $t_{\rm gas}$ using the global gas fraction,
\begin{equation}
    f_{\rm gas} = \frac{M_{\rm gas}}{M_{\rm gas} + M_\star},
\end{equation}
and assume an exponential gas-depletion law with a characteristic depletion timescale $\tau_{\rm dep} = 3\,{\rm Gyr}$, consistent with typical values for nearby star-forming disks (\citet{2008AJ....136.2782L, 2024A&A...689A.221H}). Under~these assumptions the gas fraction evolves as $f_{\rm gas} \propto \exp(-t / \tau_{\rm dep})$, which implies for the time to reach a fraction $1/e$ of $f_{\rm gas}$
\begin{equation}
    t_{\rm gas} = - \tau_{\rm dep} \ln f_{\rm gas}.
    \label{eq:tgas}
\end{equation}

We compute this quantity for all SPARC galaxies with $0 < f_{\rm gas} < 1$, and~clip the resulting ages at the Hubble time $t_H = 13.8\,{\rm Gyr}$. {All systems considered are local ($z \approx 0$); thus, we adopt the present-day Hubble time. A~redshift-dependent $t_H(z)$ would not alter the qualitative ordering.} These $t_{\rm gas}$ values should not be interpreted as physical stellar ages; they serve only as relative indicators of evolutionary state (e.g.\ young, gas-rich dwarfs versus older, gas-poor spirals). Importantly, they play no role in the computation of the dynamical quantities $D_M$, $\mathcal{T}$, $\mathcal{T}_1$ or $\mathcal{A}$. {These toy ages are used purely for visual colour-coding and do not enter any dynamical computation.}

The same SPARC catalogue is used to divide galaxies into four broad morphological surface brightness classes: HSB spirals, LSB spirals, HSB dwarfs, and~LSB dwarfs. We use the de Vaucouleurs $T$-type to distinguish dwarfs from spirals, adopting $T \ge 7$ (Sd, Sm, Im, Blue Compact Dwarf) as dwarfs and $T<7$ as spirals, and~we classify discs as LSB when $\log_{10} SB_{\rm disk} < 2.5$, corresponding to a central surface brightness of $\approx 300\,L_\odot\,{\rm pc}^{-2}$ (see \citet{2016AJ....152..157L}). Galaxies with non-finite $SB_{\rm disk}$ values are conservatively assigned to the LSB class. This four-way division is used for plotting and for visualising where different morphological families lie in the diagnostic~planes.

\begin{figure}[H]
    \includegraphics[width=0.95\linewidth]{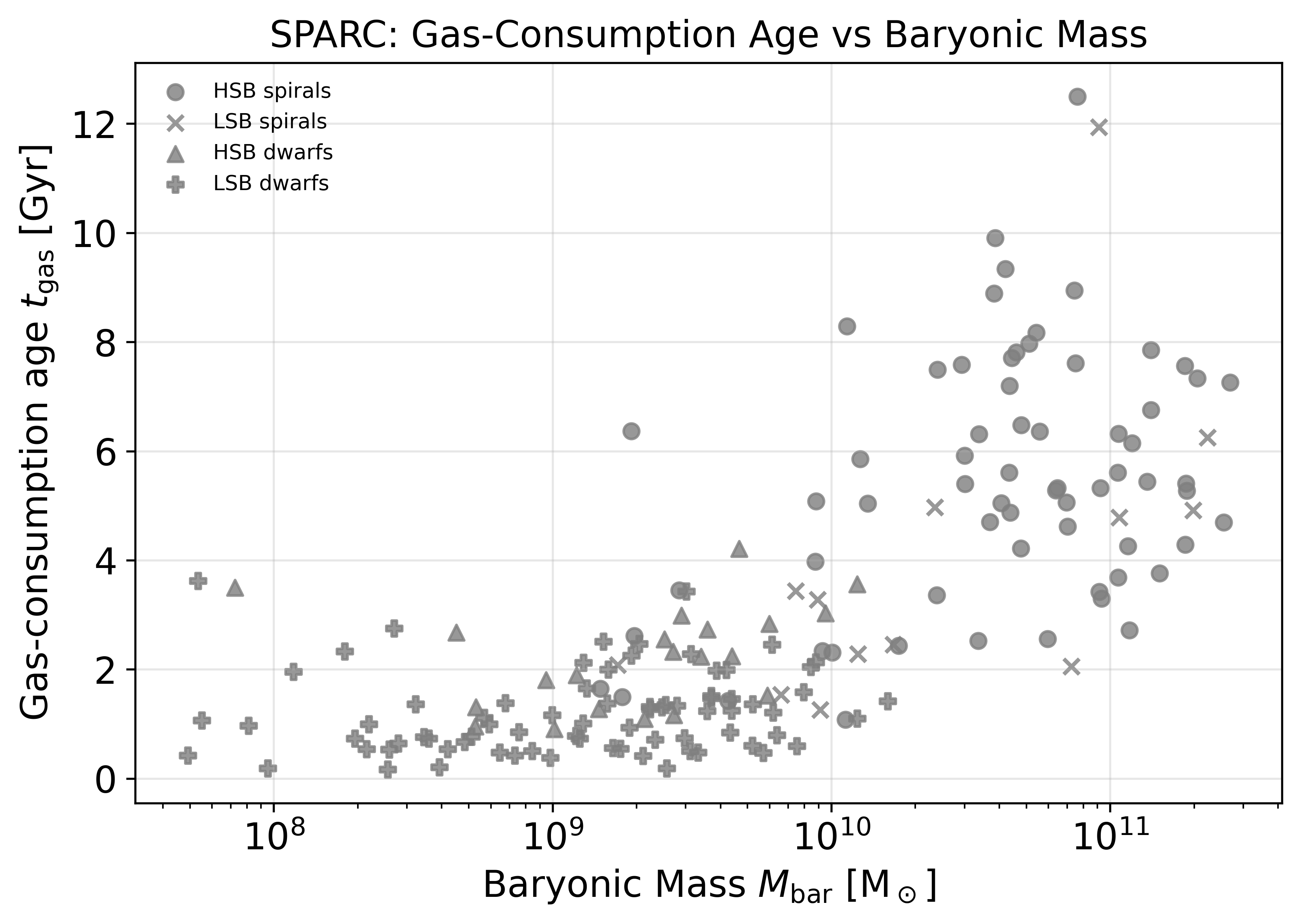}
    \caption{The gas-consumption age $t_{\rm gas}$ as a function of baryonic mass for the SPARC galaxies, estimated from the gas fraction assuming an exponential depletion law with $\tau_{\rm dep}=3\,{\rm Gyr}$. LSB dwarfs populate the youngest part of the distribution, while HSB spirals show substantially older gas-consumption ages. This behaviour reflects the extended star-formation histories and high gas fractions characteristic of diffuse, low-acceleration systems. }
    \label{fig:sparc_tgas}
\end{figure}

Figure~\ref{fig:sparc_tgas} illustrates the distribution of the gas-consumption ages, $t_{gas}$, as~a function of baryonic mass for the SPARC galaxies. Low-mass and LSB dwarfs typically have $t_{\rm gas} \leq 2\,{\rm Gyr}$, reflecting their high gas fractions and slow, prolonged star-formation histories and youth (\citet{2024A&A...689A.221H}, \citet{1997ApJ...481..689M, 2004AJ....128.2170H}, \mbox{\citet{2011ApJ...739....5W}}). HSB spirals, by~contrast, exhibit substantially older gas-consumption ages, often $6$–$12\,{\rm Gyr}$, consistent with more evolved stellar populations, higher stellar-to-gas ratios, and~earlier formation epochs (\citet{2001ApJ...550..212B, 2005MNRAS.362...41G},\mbox{\citet{2008AJ....136.2782L}}). These trends are in line with the broader picture that diffuse, low-acceleration galaxies remain chemically and dynamically unevolved, whereas massive disks form the bulk of their stars at earlier cosmic~times.

Stellar ages for ETGs are taken directly from \citet{2015MNRAS.448.3484M}, who provide luminosity-weighted ages derived from full spectral fitting. For~SPARC galaxies we rely exclusively on the gas-consumption ages described above; we do not mix these with heterogeneous stellar ages from the literature. For~MCOs we do not compute ages explicitly; instead, they are treated as old systems with typical ages $\geq 10$–$12\,{\rm Gyr}$ (\citet{2008MNRAS.386..864D}) and are shown in the diagnostic planes with a fixed symbol that is not tied to the colour-coded age~scale.

All radii, masses, and~velocities are homogenized through consistent unit conversions and, where necessary, matched definitions across the heterogeneous sample. For~each object in the combined catalogue we compute the crossing time, relaxation time, mean acceleration, MOND radius, and~all the dynamical indices introduced in this paper using the definitions in Section~\ref{sec:theory}.  The~final sample spans nearly eight orders of magnitude in baryonic mass and covers the full observed range of surface brightnesses. This broad dynamical baseline makes it ideally suited to testing whether the $D_M$ defined entirely from baryonic quantities captures the structural and evolutionary organisation of stellar systems across the entire mass spectrum. {The final combined sample contains $N_{ETGs}$ = 258 ETGs, $N_{SPARC}$ = 175 SPARC galaxies, and~$N_{MCOs}$ = 32 MCOs.}

\section{Results: Dynamical Diagnostic~Planes}
\label{sec:results}

Using the dynamical quantities introduced in Section~\ref{sec:theory}, we construct three complementary diagnostic planes that jointly characterise the collisionality, dynamical maturity, acceleration regime and MOND depth of ETGs, HSB and LSB---spirals and dwarfs and MCOs. Although~these systems span more than eight orders of magnitude in mass and exhibit diverse morphologies, they occupy coherent and physically interpretable loci in all three diagnostic planes. Taken together, these planes reveal a continuous dynamical sequence that links galaxies and star clusters through their internal accelerations, crossing times, and~positions relative to the MOND radius. Below~we discuss each diagnostic plane in turn, starting from the most fundamental dynamical division and moving toward structurally motivated MOND~quantities.

\subsection{\texorpdfstring{$\mathcal{T}_1$}{T1} vs.\ \texorpdfstring{$\mathcal{A}$}{A}}
\label{sec:results_tcross_trelax}

\begin{figure}[H]
    \includegraphics[width=0.9\linewidth]{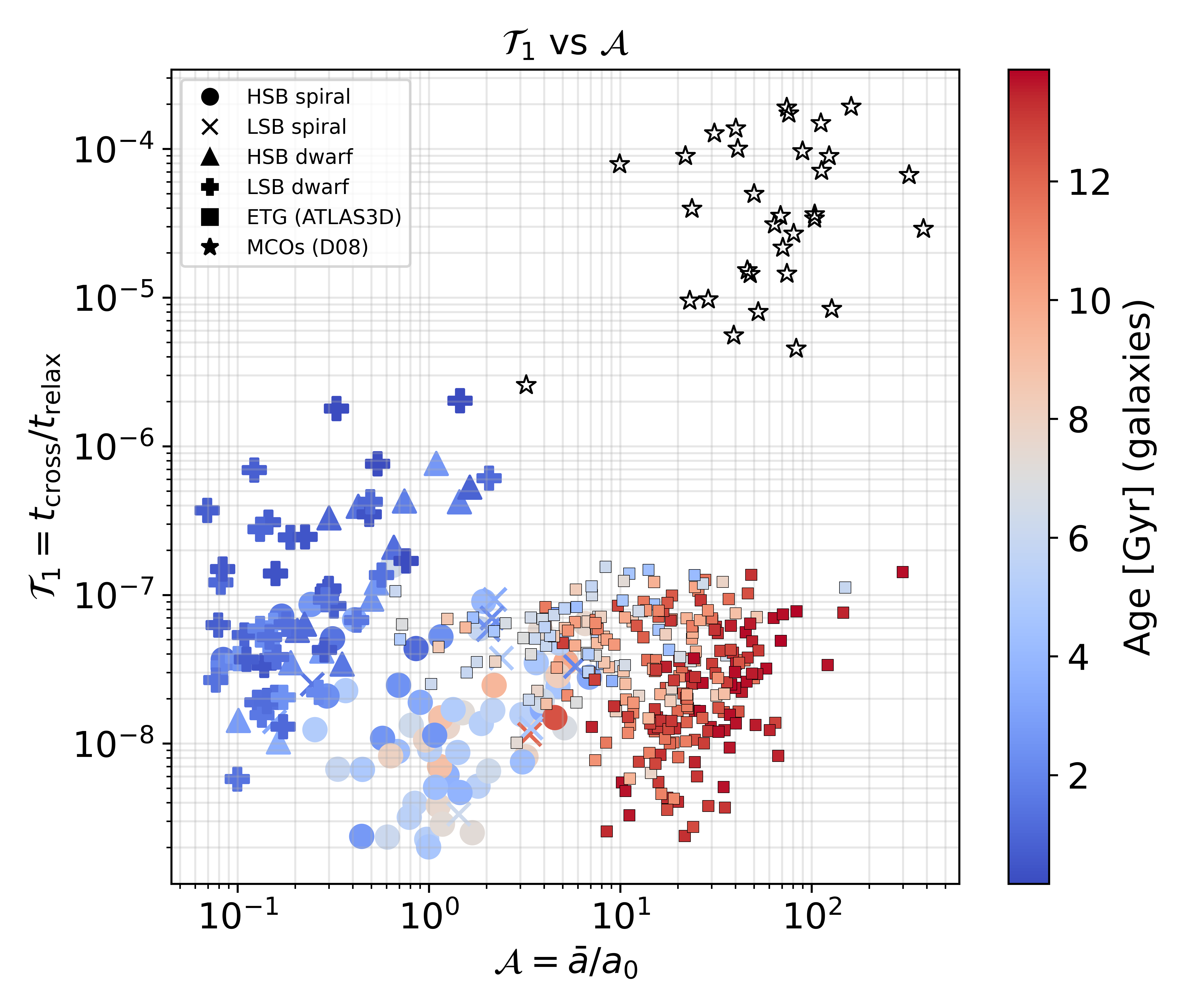}
    \caption{Collisionality index, $\mathcal{T}_1$ as a function of acceleration index, $\mathcal{A}$. The~diagnostic cleanly separates collisional stellar systems (MCOs) from collisionless galaxies. The~colour bar depicts the age ($t_{\rm age}$ values and other as defined in Section~\ref{sec:data}).}
    \label{fig:tcross_trelax}
\end{figure}

Figure~\ref{fig:tcross_trelax} compares collisionality index, $\mathcal{T}_1$ with the acceleration index, $\mathcal{A}$. Because~the relaxation time scales as $t_{\rm relax} \propto N t_{\rm cross}$, this ratio serves as a direct measure of collisionality: large values identify systems where two-body encounters drive evolution, while small values correspond to collisionless~dynamics.
\vspace{-6pt}
MCOs occupy the high-$\mathcal{T}_1$ region of the diagnostic plane, reflecting their collisional nature. More massive MCOs and compact ellipticals lie near the classical relaxation boundary $t_{\rm relax} \approx t_H$, confirming that these objects inhabit the transitional regime previously discussed in \citet{1998MNRAS.300..200K} and \citet{2008MNRAS.386..864D}. All galaxies: ETGs, HSB and LSB---spirals and dwarfs lie far below this boundary across the full range of accelerations probed. This diagnostic plane therefore recovers the fundamental physical distinction between galaxies and MCOs, and~it does so through a purely dynamical plane that naturally incorporates the MOND acceleration~scale.

\subsection{\texorpdfstring{$\mathcal{T}$}{tcross/tH} vs.\ \texorpdfstring{$\mathcal{A}$}{abar/a0}}
\label{sec:results_tcross_tH}

\begin{figure}[H]
    \includegraphics[width=0.9\linewidth]{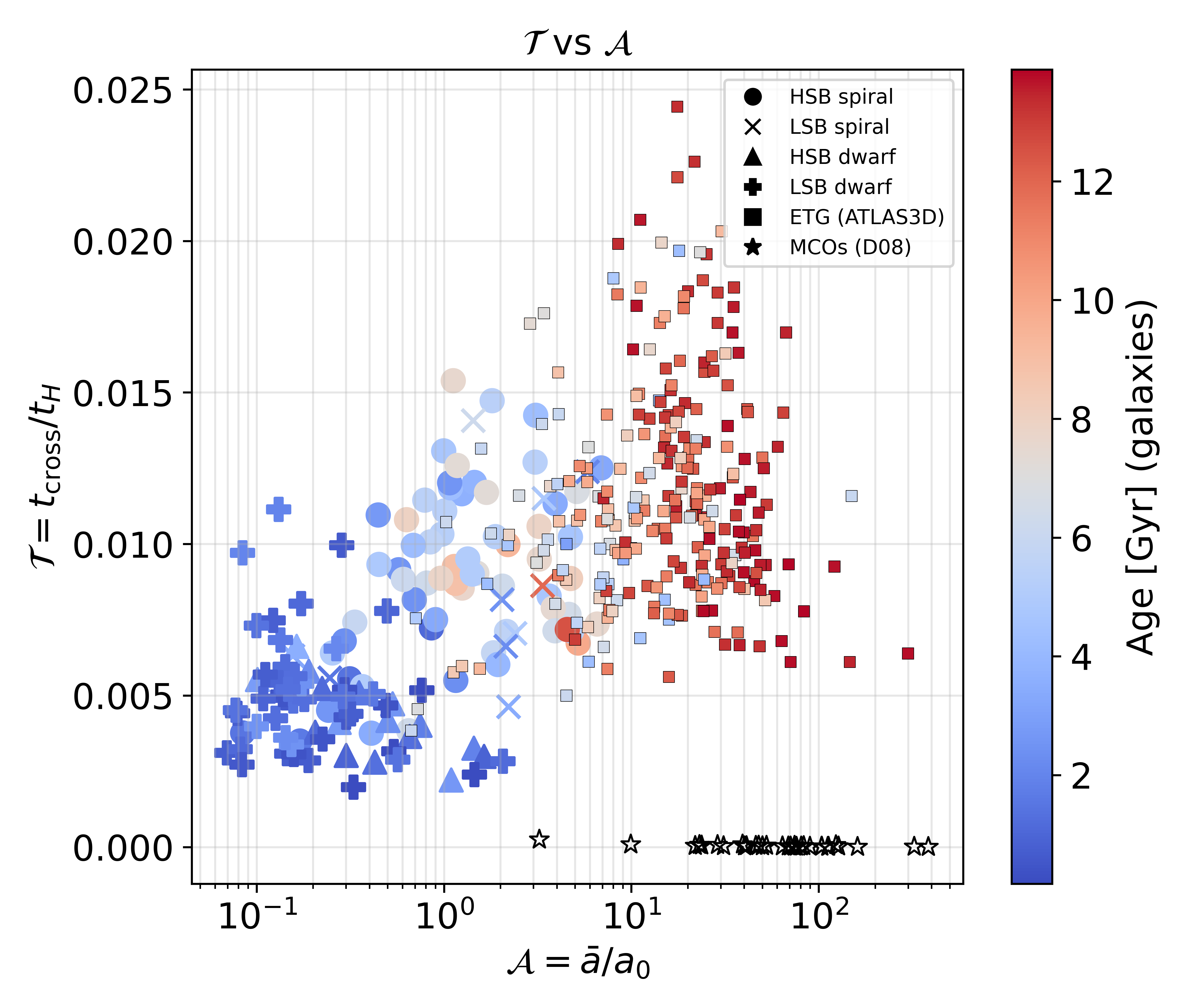}
    \caption{Dynamical maturity index, $\mathcal{T}$ as a function of $\mathcal{A}$. A~continuous acceleration–timescale sequence links all galaxy types and MCOs. The~vertical axis is shown on a linear scale to highlight the relative spread in $\mathcal{T}$ among the galaxy populations; MCOs, which occupy a very narrow range in $\mathcal{T}$, therefore appear cluttered near the lower edge of the diagnostic plane. The~colour bar is as in Figure~\ref{fig:tcross_trelax}.}

    \label{fig:tcross_tH}
\end{figure}

Figure~\ref{fig:tcross_tH} shows \(\mathcal{T}\) as a function of \(\mathcal{A}\). Unlike the relaxation time, which diverges for collisionless galaxies and therefore cannot serve as a universal dynamical age indicator, the~Hubble time provides a finite and uniform reference scale. \(\mathcal{T}\) is therefore well-defined for all systems, from~MCOs to giant~ETGs.

MCOs occupy the extreme high-acceleration regime with \(\mathcal{T} \approx 0\), reflecting their extremely short dynamical timescales. ETGs also lie at high \(\mathcal{A}\) but exhibit moderately small \(\mathcal{T}\), consistent with early formation and long-term dynamical stability. Spirals span an intermediate range in both variables, while diffuse dwarfs populate the lowest-acceleration end but do not achieve large \(\mathcal{T}\); their small sizes yield relatively short crossing times despite being deep-MOND~systems.

The combined distribution reveals that neither morphology nor total mass sets a galaxy’s dynamical age. Instead, \(\mathcal{A}\) serves as the primary organising index: self-gravitating systems follow a broad structural progression in the (\(\mathcal{T}\)--\(\mathcal{A}\)) plane, with~compactness (or internal acceleration) regulating crossing times rather than the MOND regime itself. Overall, the~(\(\mathcal{T}\)--\(\mathcal{A}\)) plane reveals a structural hierarchy rather than a monotonic evolutionary sequence: dynamical age is primarily governed by internal compactness, with~\(\mathcal{A}\) acting as the unifying index across galaxies and star clusters. This plane demonstrates that dynamical age is not determined by galaxy mass or morphology. Instead, internal acceleration (compactness) acts as the primary organising~index.

\subsection{\texorpdfstring{$D_M$}{DM} vs.\ \texorpdfstring{$\mathcal{T}$}{T}}
\label{sec:results_DM_vs_T}

\begin{figure}[H]
    \includegraphics[width=0.9\linewidth]{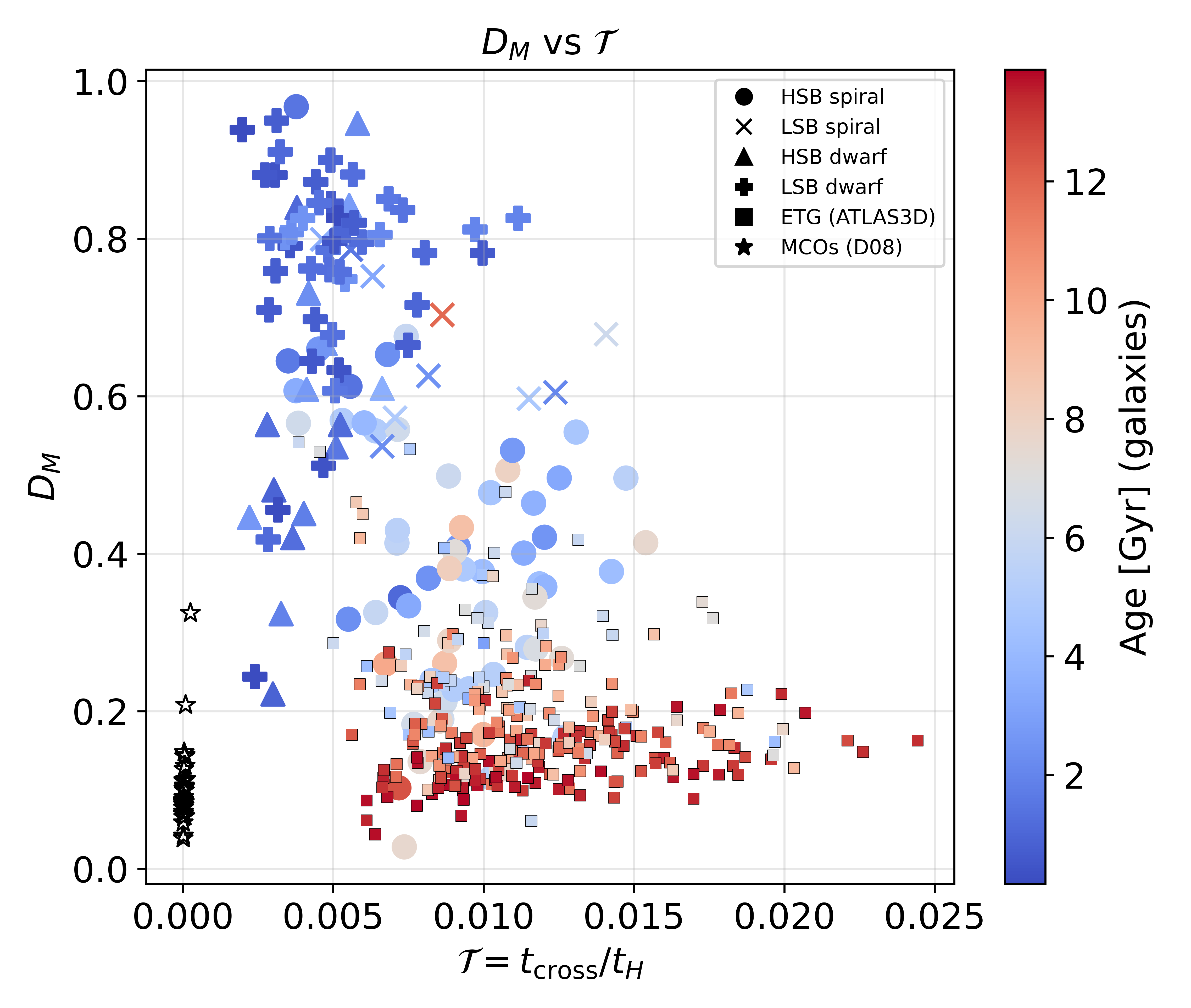}
    \caption{\textit{MOND depth index}, $D_M$ versus dynamical maturity index, $\mathcal{T}$, colour-coded by stellar population age. A~continuous sequence emerges that links structural depth, dynamical maturity, and~stellar age across all galaxy types. A~linear scale is adopted for the $\mathcal{T}$-axis to enhance the visual spread of the galaxy populations along the horizontal direction; as a consequence, the~MCOs, which occupy a very narrow range in $\mathcal{T}$, appear visually cluttered near the left edge of the diagnostic plane. The~colour bar is as in Figure~\ref{fig:tcross_trelax}.}

    \label{fig:DM_vs_T}
\end{figure}

Figure~\ref{fig:DM_vs_T} compares the $D_M$ a structural measure of how deeply a system lies inside its MOND radius with the dynamical maturity index, $\mathcal{T}$. Although~these two quantities {are derived from related but distinct definitions of dynamical depth within the MOND framework}, the~diagnostic plane formed displays a remarkably well-defined~sequence.

ETGs and MCOs occupy the region of low $D_M$ and low $\mathcal{T}$, indicating deep potential wells and dynamically old configurations. HSB spirals bridge the gap toward higher $D_M$, while LSB galaxies and diffuse dwarfs populate the opposite extreme of large $D_M$ and large $\mathcal{T}$, reflecting shallow gravitational potentials and slow dynamical evolution. ({Although diffuse dwarfs lie in the deep-MOND regime (large DM), their small physical sizes prevent $\mathcal{T}$ from reaching the extreme values seen in larger diffuse disks.})

When colour-coded by stellar population age, the~sequence becomes an evolutionary axis: dynamically deep systems are old, while dynamically shallow systems are young. Thus, $D_M$ and $\mathcal{T}$ jointly encode both the structural maturity and collapse history of stellar systems, providing a dynamical explanation for downsizing trends in galaxy~populations.

\section{Discussion}
\label{sec:discussion}

The diagnostic planes presented in Section~\ref{sec:results} demonstrate that self-gravitating stellar systems across the full observable hierarchy from diffuse dwarfs and LSB disks to massive ETGs, HSB spirals and MCOs populate a remarkably continuous and physically interpretable set of sequences. Despite spanning eight orders of magnitude in baryonic mass, and~exhibiting a wide range of structural and kinematic properties, these systems align in a manner that is naturally organised by $D_M$, $\mathcal{T}$, $\mathcal{A}$, and~$\mathcal{T}_1$. Taken together, these relations define an empirical dynamical backbone that links structural depth, collapse history, and~stellar age across all classes of stellar systems. The~continuous behaviour seen in our MOND-based diagnostic planes reflects the structural continuity long recognised between MCOs (which includes massive globular clusters and UCDs) and dwarf galaxies~\citep{2011MNRAS.414.3699M}. This supports the view that these systems trace a single underlying sequence rather than representing sharply distinct~classes.

A central result is that galaxies and MCOs form a unified dynamical sequence when expressed in the $(D_M,\,\mathcal{T})$ plane. Low-$D_M$ systems (ETGs and MCOs) exhibit short crossing times, high accelerations, and~old stellar populations, consistent with early and rapid collapse (e.g. \citet{2010ApJ...709.1018V, 2005ApJ...621..673T, 2015MNRAS.448.3484M, 2019A&A...632A.110Y, 2022MNRAS.516.1081E}). At~the opposite extreme, diffuse dwarfs and LSB spirals reside at large $D_M$ and large $\mathcal{T}$, reflecting their shallow potentials, slow collapse, and~young or extended star-formation histories (e.g. \citet{2024A&A...689A.221H}). HSB spirals bridge these two populations with intermediate MOND depths and dynamical ages, forming the central spine of the overall~sequence.

The colour-coding by stellar population age further shows that this dynamical sequence is simultaneously an evolutionary sequence: dynamically deep systems are predominantly old, while dynamically shallow systems are young. In~this sense, the~$D_M$ provides a physical underpinning for the well-established phenomenon of downsizing (\citet{1996AJ....112..839C, 2005ApJ...621..673T, 2019A&A...632A.110Y, 2022MNRAS.516.1081E}), long recognised in observational studies of galaxy evolution. Within~the MOND framework, this behaviour emerges naturally from the scaling of the MOND radius $r_M \propto M_{\rm bar}^{1/2}$ and from the sensitivity of collapse times to internal acceleration: massive galaxies collapse within $r_M$, achieving high accelerations and quickly quenching, whereas low-mass, diffuse systems form and evolve in the deep-MOND regime with prolonged dynamical and star-forming timescales (\citet{2008MNRAS.386.1588S, 2022MNRAS.517.3613K, 2022Symm...14.1331B}). {In standard dark-matter based models, the~relation between baryonic structure and total dynamical depth depends on halo concentration, assembly history, and~baryon–halo coupling, which introduce additional degrees of freedom and scatter (e.g. \citet{2008Natur.455.1082D}).}

{In addition to dark-matter based and MOND frameworks, a~number of purely Newtonian models that explicitly account for realistic galaxy geometry (e.g., disk or oblate mass distributions) have been explored in the literature (e.g., \citet{2020Galax...8....9F}, \mbox{\citet{2020Galax...8...54H, 2020Galax...8...47H}}). These studies consider whether detailed Newtonian force balance using baryonic mass distributions alone can reproduce aspects of observed rotation curves, and~we include this acknowledgment here for completeness.}

The $\mathcal{T}_1$ versus $\mathcal{A}$ plane recovers, with~modern data, the~classical collisionless–collisional divide originally emphasised by \citet{1998MNRAS.300..200K} and extended in \citet{2008MNRAS.386..864D}. MCOs occupy the collisional regime, while all galaxies irrespective of morphology or mass lie securely within the collisionless domain. The~addition of the MOND acceleration axis sharpens this boundary by showing that high-acceleration, compact stellar systems approach the relaxation threshold from above, whereas diffuse galaxies remain far below it. {ETGs likely reach quasi-equilibrium via rapid collective processes such as violent relaxation and dissipative gas collapse, rather than via two-body relaxation.}

Several systematic uncertainties must be acknowledged. The~definition of the characteristic radius differs somewhat across ETGs, spirals, and~dwarfs; while these choices follow standard practice (e.g., \citet{2011MNRAS.413..813C, 2016AJ....152..157L}), they introduce scatter in both $\bar{a}$ and $t_{\rm cross}$. Stellar ages for gas-rich dwarfs are uncertain and often spatially variable, which may blur the age–dynamics correspondence at low mass. The~computation of $M(<r_M)$ relies on Hernquist or exponential approximations that, while reasonable, cannot capture all structural diversity, especially for disturbed or composite systems. Finally, some MCOs may host massive central black holes (\citet{2014Natur.513..398S}) or sub-clusters of stellar-mass black holes (\citet{2021MNRAS.502.5185M}), which could influence their internal accelerations. Yet despite these uncertainties, the~observed dynamical sequences remain strikingly well defined, suggesting that the key physical trends are~robust.

The dynamical diagnostic planes introduced in this work occupy a conceptual role somewhat analogous to that of the Hertzsprung–Russell (HR) diagram for stars (\citet{1911AN....190..119H, 1914PA.....22..331R}). The~classical HR diagram provides a unified framework in which stars populate distinct, physically interpretable sequences whose positions reflect their internal structure and evolutionary state. In~an analogous, though~not identical, manner our diagnostic planes show that galaxies and stellar clusters do not populate the dynamical indices space randomly: HSB and LSB spirals and dwarfs, ETGs and MCOs all occupy well-defined and contiguous regions in the planes defined by $D_M$, $\mathcal{T}$, $\mathcal{T}_1$, and~$\mathcal{A}$. The~location of a system in these planes encodes its dynamical maturity, internal acceleration regime, compactness, and~collisional state, thereby offering a unified classification scheme across the full hierarchy of stellar systems. {Extending this framework to galaxy groups and clusters would provide an important test, particularly given the known residual mass discrepancy in MOND at cluster scales has recently been amended \citep{2026PhRvD.113d3027Z}.}

The emergence of continuous sequences from deep-MOND, dynamically old ETGs to diffuse, dynamically young LSB disks, and~further to relaxation-dominated MCOs demonstrates that these planes provide a {useful} structural–dynamical analogue to the HR diagram. They thus offer a new way to characterise galaxies in terms of their dynamical state and evolutionary maturity within the MOND framework. Overall, the~$D_M$ and $\mathcal{T}$ together provide a simple, baryon-based, and~physically grounded framework for understanding the internal dynamics, structural maturity, and~evolutionary pathways of stellar systems. The~continuity of the sequences uncovered here indicates that galaxies and MCOs form a dynamically ordered family rather than a set of disjoint morphological categories. {In this sense, $D_M$ and $\mathcal{T}$ together define a physically motivated dynamical classification framework that organises stellar systems across a wide range of mass and~morphology.}
\section{Conclusions}
\label{sec:conclusions}

In this work we have introduced a new, physically motivated dynamical index, $D_M$, defined purely from the baryonic mass distribution and the universal MOND acceleration scale $a_0$. Together with $\mathcal{T}$, $\mathcal{T}_1$ and $\mathcal{A}$, these indices provide a compact description of dynamical depth, collapse history, and~internal gravitational regime for stellar systems across the full mass~spectrum.

{Using a combined sample of ETGs from ATLAS\textsuperscript{3D}, spirals and dwarfs from SPARC and compact stellar systems from \citet{2008MNRAS.386..864D}, we have shown that stellar systems occupy a continuous dynamical sequence in the primary diagnostic planes defined by $(D_M,\mathcal{T})$, $(\mathcal{T},\mathcal{A})$, and~$(\mathcal{T}_1,\mathcal{A})$} (A supplementary structural trend between $D_M$ and total baryonic mass is presented in Appendix~\ref{appendix:DM_vs_M}). Dynamically deep systems with small $D_M$ (ETGs and compact stellar systems) are universally old, while dynamically shallow systems with high $D_M$ (LSB disks and diffuse dwarfs) are young and gas-rich. {This provides a baryonic dynamical framework consistent with the observed downsizing phenomenon} (\citet{1996AJ....112..839C, 2005ApJ...621..673T, 2010MNRAS.404.1775T, 2015MNRAS.448.3484M, 2019A&A...632A.110Y}), linking stellar-population age directly to $D_M$, $\mathcal{T}$ and $\mathcal{A}$.

The $(\mathcal{T}_1,\mathcal{A})$ plane naturally recovers the classical ``galaxy versus star-cluster'' boundary proposed by \citet{1998MNRAS.300..200K} and extended by \citet{2008MNRAS.386..864D}: all galaxies occupy the collisionless regime with $t_{\rm relax} > t_H$, while MCOs lie in the collisional regime. 
{The MOND radius $r_M$ provides a physically meaningful collapse scale: systems with most of their baryons inside $r_M$ tend to exhibit short dynamical times and old stellar populations, whereas systems with most of their baryons outside $r_M$ evolve more slowly in the deep-MOND regime.}

Taken together, these findings demonstrate that $D_M$, $\mathcal{T}$, $\mathcal{T}_1$  and $\mathcal{A}$ provide a powerful and unified framework for understanding structural evolution, dynamical evolution, and~stellar-population age across the full spectrum of stellar systems. 
{The coherence of the resulting sequences, and~their connection to stellar ages, illustrate how the MOND framework provides a natural context for organising structural and dynamical trends across stellar systems. }Future work should extend this analysis to larger and higher-redshift samples, including JWST galaxies, ultra-diffuse systems in diverse environments, and~massive compact objects, in~order to test the robustness and universality of the $D_M$-based dynamical classification proposed here.




\vspace{6pt}
\authorcontributions{
R.E. developed the initial concept of the MOND depth index and performed all calculations and data analysis. P.K. contributed to the development of the dynamical-age indicator $\mathcal{T}$ and advised on the connection between the acceleration ratio and relaxation time. R.E. prepared the first draft of the manuscript; R.E. and P.K. jointly revised and developed the manuscript. All authors have read and agreed to the published version of the manuscript.
}

\funding{This work is not part of a funded~project.} 

\dataavailability{The original contributions presented in this study are included in the article/supplementary material. Further inquiries can be directed to the corresponding author(s).} 

\acknowledgments{The 
 author R.E. would like to thank KFPP for their support. P.K. would like to thank the DAAD Bonn---Eastern European Exchange program at the University of Bonn and Charles University in Prague for support. The~authors thank J. Dabringhausen for suggesting the connection between the acceleration ratio and the relaxation-timescale condition for collisional systems, which helped refine the interpretation of the dynamical~sequence.}

\conflictsofinterest{The authors declare no conflicts of~interest.} 

\appendixtitles{yes} 
\appendixstart
\appendix
\section[\appendixname~\thesection]{Supplementary Structural Trend: $\boldsymbol{D_M}$ versus Baryonic~Mass}
\label{appendix:DM_vs_M}

\begin{figure}[H]
    \includegraphics[width=0.9\linewidth]{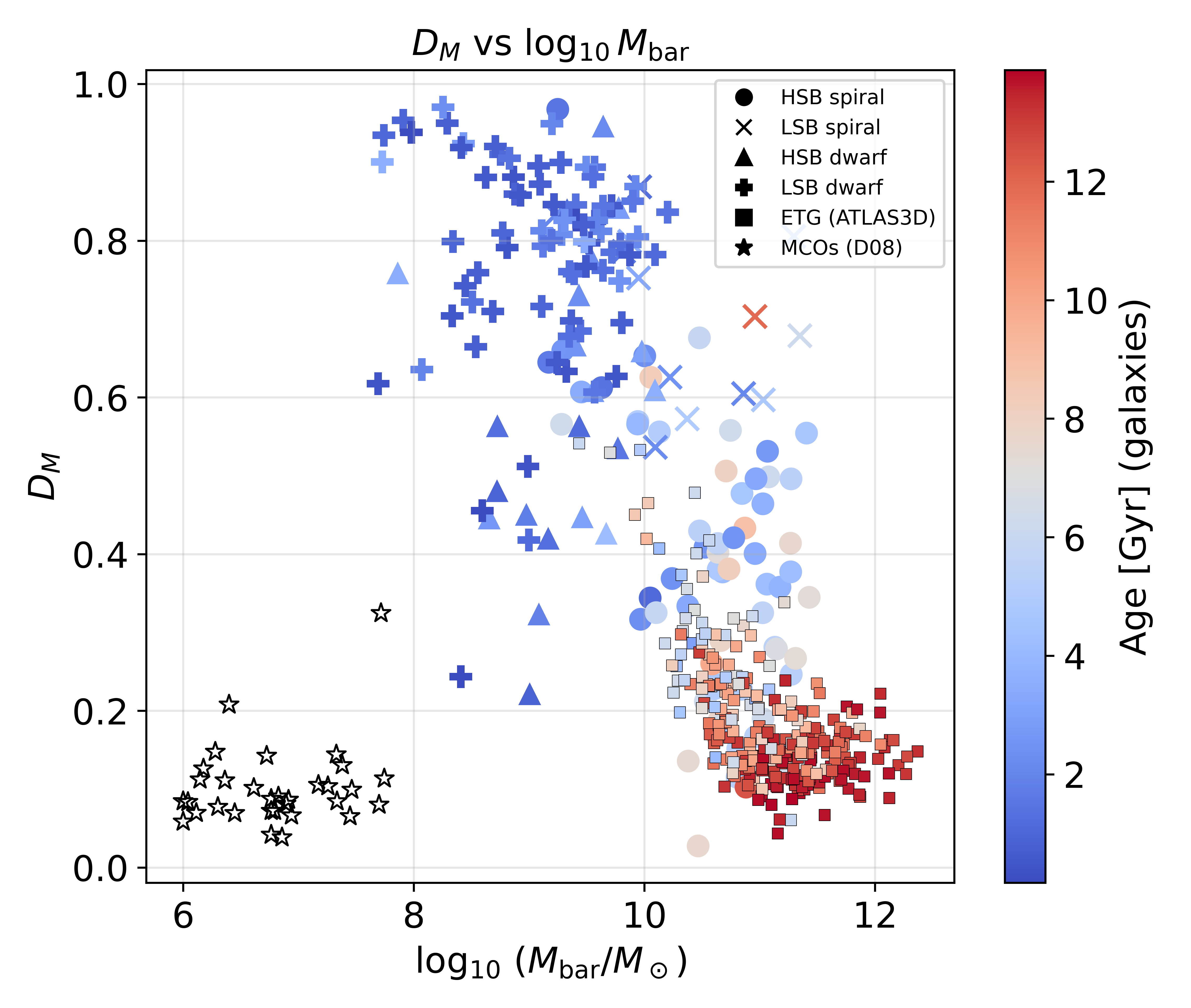}
    \caption{{MOND} 
 depth index $D_M$ as a function of total baryonic mass.
    The relation reflects the scaling of the MOND radius $r_M \propto M_{\rm bar}^{1/2}$,
    shown here for completeness as a supplementary structural trend within the sample. The~colour bar is as in Figure~\ref{fig:tcross_trelax}.}
    \label{fig:DM_vs_M}
\end{figure}

For completeness, we illustrate in Figure~\ref{fig:DM_vs_M} the empirical relation between
the MOND depth index $D_M$ and the total baryonic mass of the systems~considered.
\vspace{-6pt}
Low-mass dwarf systems tend to occupy the regime $D_M \approx 1$, indicating that a large
fraction of their baryons lie outside their MOND radius. Intermediate-mass spiral galaxies
populate a transitional region, while early-type galaxies and compact stellar systems
generally exhibit $D_M \ll 1$, consistent with higher internal accelerations and greater
baryonic concentration within $r_M$.

Since the MOND radius scales as $r_M \propto M_{\rm bar}^{1/2}$, a~systematic variation of
$D_M$ with mass is expected at the level of dimensional scaling. We emphasise, however,
that this relation should not be interpreted as a strict mass threshold or universal
boundary. Known extended low-surface-brightness systems and galaxy groups demonstrate
that structural depth depends not only on total mass but also on the spatial distribution
of~baryons.

We therefore present this plane as a supplementary structural trend within the present
dataset, rather than as a primary diagnostic. The~main results of this work remain
the dynamical classification planes involving $\mathcal{A}$, $\mathcal{T}$, and~$D_M$ discussed in
Section~\ref{sec:results}.

\begin{adjustwidth}{-\extralength}{0cm}

\reftitle{References}




\PublishersNote{}
\end{adjustwidth}
\end{document}